# On the noise-resolution duality, Heisenberg uncertainty and Shannon's information

T.E. Gureyev[1,2,3], F. de Hoog[1], Ya.I. Nesterets[1,3] and D.M. Paganin[2]

[1] *Commonwealth Scientific and Industrial Research Organisation, Clayton, VIC 3168, Australia;*
[2] *Monash University, Clayton, VIC 3800, Australia;*
[3] *University of New England, Armidale, NSW 2358, Australia.*

15 March 2015

## Abstract
Several variations of the Heisenberg uncertainty inequality are derived on the basis of "noise-resolution duality" recently proposed by the authors. The same approach leads to a related inequality that provides an upper limit for the information capacity of imaging systems in terms of the number of imaging quanta (particles) used in the experiment. These results can be useful in the context of biomedical imaging constrained by the radiation dose delivered to the sample, or in imaging (e.g. astronomical) problems under "low light" conditions.

## 1. Introduction

Among the most important characteristics of many imaging, scattering and measuring experimental setups (systems) are the spatial resolution and the signal-to-noise ratio (SNR) [1, 2]. For mainly historical reasons (abundance of photons in typical visible light imaging applications), the two properties are usually considered separately, even though any applied physicists or optical engineer would be aware of an intrinsic link between them. These two characteristics, and the interplay between them, have attained additional relevance in recent years in the context of biomedical imaging, where the samples are sensitive to the radiation doses [3], in certain astronomical methods where the detectable photon flux can be extremely low [4], as well as in some other problems, including those related to foundations of quantum physics [5]. In X-ray medical imaging, for example, it is critically important to minimize the radiation dose delivered to the patient, while still being able to obtain 2D or 3D images with sufficient spatial resolution and SNR in order to detect the features of interest, such as small tumours [6, 7]. In this context, an imaging system (such as e.g. a CT scanner) must be able to maximize the amount of relevant information that can be extracted from the collected images, while keeping sufficiently low the number of X-ray photons impinging on the patient. The present paper addresses some mathematical properties of generic imaging systems that are likely to be important in the context of designing the next generation of medical imaging instruments, and may also have relevance to some fundamental aspects of quantum physics and information theory.

We have recently introduced [6, 7] a dimensionless "intrinsic quality" characteristic $Q_S$ which incorporates both the noise propagation and the spatial resolution properties of a linear shift-invariant (imaging) system:

$$Q_S = \frac{SNR_{out}}{F_{in}^{1/2} (\Delta x)^{n/2}}, \tag{1}$$

where $n$ is the dimensionality of the input data ($n = 2$ corresponds to conventional planar images), $F_{in}$ is the mean value of the incident particle/quanta fluence (the number of incident particles per $n$-dimensional volume), $\Delta x$ is the spatial resolution of the imaging system and $SNR_{out} = S_{out} / \sigma_{out}$ is the output signal-to-noise ratio. Here $S_{out} = \int F_{out}(\mathbf{x}) d\mathbf{x}$ is the output signal, $F_{out}(\mathbf{x})$ is the output fluence at point $\mathbf{x}$ and $\sigma_{out}$ is the standard deviation of noise in the output signal. Here it is assumed that the incident particle density is a spatially stationary random process, with a mean value (the same at all points of the "entrance aperture") equal to $F_{in} = N_q / A^n$, where $N_q$ is the total number of incident quanta and $A^n$ is the "area" of the entrance aperture of the imaging system. Note that, because $Q_S$ is normalised with respect to the incident fluence, it may be viewed as "imaging quality per single incident particle". In practice, if the incident fluence rate or the exposure time can be increased, the quality of the resultant image would be expected to increase too (normally, in proportion to $F_{in}^{1/2}$). However, in applications where the imaging quanta are at premium (e.g. in low-light imaging) or where the irradiation dose delivered to the sample is critical (as in X-ray or electron imaging of biological samples), $Q_S$ represents a key performance indicator of the imaging system.

We have previously shown [6, 8, 9] that when the total number of imaging quanta is fixed, a duality exists between the signal-to-noise and the spatial resolution of the imaging system and, as a result, the intrinsic quality $Q_S$ has an absolute upper limit (maximum):

$$Q_S^2 \leq 1 / C_n, \tag{2}$$

where $C_n = 2^n \Gamma(n/2) n(n+2) / (n+4)^{n/2+1}$ is the Epanechnikov constant [10, 6]. More precisely, it was shown in [8] that inequality (2) holds and is exact for linear shift-invariant (LSI) systems (see the precise definition in the next section) with a point-spread function (PSF) $T(\mathbf{x})$ having finite mathematical expectation, variance and energy, and the maximum is achieved on Epanechnikov PSFs $T_E(\mathbf{x}) = (1 - |\mathbf{x}|^2)_+$, where the subscript "+" denotes that $T_E(\mathbf{x}) = 0$ at points where the expression in brackets is negative. Further details about this result are given in the next section.

Although the definition of the intrinsic quality was originally introduced for LSI systems [6], later we extended it to some non-linear systems. One such example, studied in [9], corresponded to the famous Young double-slit diffraction experiment. In that context it is convenient to rewrite eq.(1) in the following equivalent form:

$$Q_S^2 = \frac{SNR_{out}^2}{N_q} \frac{A^n}{(\Delta x)^n}. \tag{3}$$

In the case of Young double-slit diffraction experiment, we used a definition of $SNR_{out}$ corresponding to the so-called "ideal observer SNR" [2], which quantified the distinguishability of the image of two identical slits of width $b = A/2$ separated by distance $d = \Delta x$ from the image of one slit with the same width located in the middle position. The number of particles, $N_q$, forming each of the two images was assumed to be the same. Obviously, the issue of distinguishability of such images is closely related to Rayleigh criterion of spatial resolution [1]. It was shown in [9] that, for any fixed number $N_q$ of image-forming quanta, the intrinsic quality, defined in eq.(3), reaches its maximum at the slit separation distance $d$ equal to $2b$, i.e. when $\Delta x = A$. In other words, the number of imaging quanta required to reliably (e.g. with $SNR_{out} \geq 5$) distinguish an image of two identical slits from the corresponding image of one slit, reaches it minimum when $d = 2b$.

In the next two sections of the present paper we will investigate the relationship between inequality (2) and the Heisenberg uncertainty inequality [11]. In Section 4, we will outline a possible link between the noise-resolution uncertainty (2) and the notion of information capacity of communication and imaging systems, as introduced by Shannon [12]. The main results will be summarized in a brief Conclusions section.

## 2. Relationship between Heisenberg and noise-resolution uncertainties

The spatial resolution $\Delta x$ of an LSI system described by the equation

$$I_{out}(\mathbf{x}) = \int T(\mathbf{x} - \mathbf{y}) I_{in}(\mathbf{y}) d\mathbf{y}, \quad \mathbf{x}, \mathbf{y} \in \mathbb{R}^n, \tag{4}$$

can be defined in terms of the width $\Delta x$ of its PSF $T(\mathbf{x})$ e.g. as follows:

$$(\Delta x)^2 = \frac{4\pi}{n} \frac{\int |\mathbf{x}|^2 T(\mathbf{x}) d\mathbf{x}}{\int T(\mathbf{x}) d\mathbf{x}} = \frac{4\pi}{n} \frac{\|\mathbf{x}t\|_2^2}{\|t\|_2^2}, \tag{5}$$

where $T(\mathbf{x}) = |t(\mathbf{x})|^2 \ge 0$ is a non-negative function with finite $L_1$ and $L_2$ norms, finite variance and zero expectation. In particular, $\int \mathbf{x} T(\mathbf{x}) d\mathbf{x} = 0$, and $\|T\|_1 = \int T(\mathbf{x}) d\mathbf{x} = \int |t(\mathbf{x})|^2 \, d\mathbf{x} = \|t\|_2^2 = \|\hat{t}\|_2^2 < \infty$, where the overhead hat symbol denotes Fourier transform, $\hat{f}(\mathbf{u}) = \int \exp(-i2\pi\mathbf{u} \cdot \mathbf{x}) f(\mathbf{x}) d\mathbf{x}$.

We also define the "angular" (or "momentum") resolution as

$$(\Delta u)^2 = \frac{4\pi}{n \|\hat{t}\|_2^2} \int |\mathbf{u}|^2 |\hat{t}(\mathbf{u})|^2 \, d\mathbf{u} = \frac{4\pi}{n} \frac{\|\mathbf{u}\hat{t}\|_2^2}{\|t\|_2^2}. \tag{6}$$

Then the Heisenberg uncertainty inequality [11] states that

$$\Delta x \, \Delta u \ge 1. \tag{7}$$

Note that the momentum of a mono-energetic plane-wave photon is equal to $\mathbf{p} = \hbar \mathbf{k}$, where $\mathbf{k}$ is the wave vector, $\hbar = h/(2\pi)$ and $h$ is the Planck constant. Identifying $\mathbf{k} \equiv 2\pi\mathbf{u}$ and $\Delta p \equiv \hbar \Delta k = h \Delta u$, inequality (7) can be written in a more conventional form:

$$\Delta x \, \Delta p \ge h. \tag{8}$$

The absence of the usual factor $1/(4\pi)$ on the right-hand side of the last inequality is due to the normalization factor $4\pi/n$ included in eqs.(5) and (6). We chose such normalization because in the imaging context it leads to a more natural scaling of the width of PSF [6]: for example, for a rectangular PSF with the side length equal to $A$, we get $\Delta x = A\sqrt{\pi/3}$ in any $\mathbb{R}^n$, according to eq.(5).

On the other hand, the noise-resolution uncertainty inequality (2) implies:

$$(\Delta x)^n \ge C_n F_{in}^{-1} SNR_{out}^2. \tag{9}$$

In order to compare this result with the Heisenberg uncertainty principle (7), we need also an analogue of eq.(9) for $\Delta u$ that would correspond to eq.(6).

According to the well-known properties of LSI systems [2], the $SNR_{out}$ from eq.(1) can in this case be expressed as follows:

$$SNR_{out} = \frac{S_{out}}{\sigma_{out}} = \frac{\iint F_{in}(\mathbf{y}) T(\mathbf{x} - \mathbf{y}) d\mathbf{y} d\mathbf{x}}{\left( A^{2n} \int W_{in}(\mathbf{u}) \, |\hat{T}(\mathbf{u})|^2 \, d\mathbf{u} \right)^{1/2}} \ , \qquad (10)$$

where $W_{in}(\mathbf{u})$ is the power spectral density of noise in the input signal. Assuming that the incident fluence is spatially stationary over the entrance aperture of area $A^n$, that it satisfies Poisson statistics and is spatially uncorrelated, we obtain: $\sigma_{in}^2 = N_q = A^{2n} \int W_{in} d\mathbf{u} = W_{in} A^n$ and hence

$$SNR_{out}^2 = \frac{N_q \, \|T\|_1^2}{A^n \, \|T\|_2^2} = \frac{F_{in} \, \|t\|_2^4}{\|t\|_4^4} . \qquad (11)$$

Substituting this into eq.(9), we obtain:

$$(\Delta x)^n \geq C_n \, \|t\|_2^4 \, / \, \|t\|_4^4 , \qquad (12)$$

where $\Delta x$ is expressed by eq.(5).

A similar "noise-resolution uncertainty" inequality can now be written for $\Delta u$ (as defined in eq.(6)) by replacing $t(x)$ with $\hat{t}(u)$ in (12):

$$(\Delta u)^n \geq C_n \, \|t\|_2^4 \, / \, \|\hat{t}\|_4^4 . \qquad (13)$$

Multiplying (12) and (13) gives us an inequality similar to the Heisenberg uncertainty (7):

$$V[t](\Delta x \, \Delta u)^n \geq C_n^2 , \qquad (14)$$

where the dimensionless quantity

$$V[t] = \|t\|_4^4 \|\hat{t}\|_4^4 \, / \, \|t\|_2^8 \qquad (15)$$

represents a kind of a "phase-space noise-to-signal ratio" (normalized with respect to the incident fluence) which characterizes a particular imaging (measuring) system.

The functional $V[t]$ is bi-invariant with respect to scaling of its argument, i.e. $V[at(b\mathbf{x})] = V[t(\mathbf{x})]$ for any positive constants $a$ and $b$, hence it does not depend on the "height" or "width" of the function $t(\mathbf{x})$, but only on its functional form. For Gaussian functions $t_G(\mathbf{x}) = a\sqrt{2\pi} \exp(-|\mathbf{x}|^2 / (2b)]$, one always has $V[t_G] = 1$. In this case, inequality (14) is weaker than (7), since the Epanechnikov constants $C_n$ are slightly smaller than 1 (for example, e.g. $C_1 = 6\sqrt{\pi/125} \cong 0.95$, $C_2 = 8/9$ and $C_3 = 60\sqrt{\pi}/7^{5/2} \cong 0.82$). It can be shown (see Appendix) that the functional $V[t]$ can be arbitrary close to zero for some functions $t(\mathbf{x})$ and can be arbitrary large for other functions. The former means that for some functions $t(\mathbf{x})$ inequality (14) gives a stronger estimate (higher lower bound) than the Heisenberg uncertainty (7).

## 3. "Incoherent" version of Heisenberg uncertainty inequality

Let us define an alternative ("incoherent") angular resolution in the following way:

$$(\tilde{\Delta}u)^2 = \frac{4\pi}{n \parallel \hat{T} \parallel_1} \int |\mathbf{u}|^2 |\hat{T}(\mathbf{u})| \, d\mathbf{u},$$ (16)

i.e. it is equal to the width of the modulation transfer function (MTF), $|\hat{T}(\mathbf{u})|$. We also introduce a new SNR, in the same way as in eq.(11), but with $|\hat{T}(\mathbf{u})|$ in place of $T(\mathbf{x})$, i.e.

$$\widetilde{SNR}_{out}^2 = \frac{F_{in} \parallel \hat{T} \parallel_1^2}{\parallel T \parallel_2^2}.$$ (17)

Then an analogue of (9) for $|\hat{T}(\mathbf{u})|$ in place of $T(\mathbf{x})$ is:

$$(\tilde{\Delta}u)^n \geq C_n F_{in}^{-1} \widetilde{SNR}_{out}^2.$$ (18)

Multiplying (9) and (18), we obtain:

$$(\Delta x \tilde{\Delta}u)^n \geq C_n^2 \parallel T \parallel_1^2 \parallel \hat{T} \parallel_1^2 / \parallel T \parallel_2^4.$$ (19)

It is easy to show that $\parallel T \parallel_1^2 \parallel \hat{T} \parallel_1^2 / \parallel T \parallel_2^4 \geq 1$. Indeed,

$$\int T^2(x)dx = \int T(x) \int \exp(i2\pi xu)\hat{T}(u)dudx \leq \int |T(x)| \, dx \int |\hat{T}(u)| \, du.$$

Therefore,

$$\Delta x \tilde{\Delta}u \geq C_n^{2/n}.$$ (20)

This can be viewed as an alternative ("incoherent") form of the Heisenberg uncertainty principle.

Equation (20) can be re-written as

$$4\pi^2 \frac{\int |\mathbf{x}|^2 |T(\mathbf{x})| d\mathbf{x}}{\int |T(\mathbf{x})| d\mathbf{x}} \frac{\int |\mathbf{u}|^2 |\hat{T}(\mathbf{u})| d\mathbf{u}}{\int |\hat{T}(\mathbf{u})| d\mathbf{u}} \geq \frac{n^2}{4} C_n^{4/n}.$$  (21)

The optimal (sharp) lower bound for the left-hand side of eq.(21) in 1D case is known as the Laue constant, $\lambda_0$ see e.g. [13]. It has been proven (see e.g. [14]) that $0.543 < \lambda_0 < 0.85024$. The constant $C_1^4/4 \cong 0.205$ is obviously much lower than the optimal bound, although strictly speaking the Laue constant is an optimal lower bound only for symmetric 1D functions $T(x)$ [13].

## 4. Relationship between noise-resolution uncertainty and Shannon's information capacity

Another uncertainty relationship can be obtained for a (broad) class of imaging (or measuring) systems with the (output) spatial resolution not exceeding the size of the entrance aperture, i.e. $\Delta x \leq A$. Multiplying both sides of eq.(9) by $(\Delta u)^n$, we obtain

$$(\Delta x \Delta u)^n \geq C_n F_{in}^{-1} (\Delta u)^n SNR_{out}^2.$$  (22)

As $F_{in}^{-1} = A^n / N_q$, $F_{in}^{-1}(\Delta u)^n = A^n (\Delta u)^n / N_q \geq 1 / N_q$, because $A\Delta u \geq \Delta x \Delta u \geq 1$. Then

$$(\Delta x \Delta u)^n \geq C_n SNR_{out}^2 / N_q.$$  (23)

In one limit case, when all output quanta are collected in a single "detector pixel" with the Poisson statistics, we have $SNR_{out}^2 / N_q = 1$. In this case inequality (23) gives only a slightly smaller lower limit for its left-hand side than the conventional Heisenberg uncertainty (7). At the other limit, when the output signal is uniformly spread over multiple "pixels" (which corresponds to narrow PSFs with $A^n \|T\|_2^2 >> \|T\|_1^2$ in eq.(11)), $SNR_{out}$ can be close to 1, even for large $N_q$, and hence $SNR^2 / N_{qnt} \sim 1 / N_{qnt}$, indicating that the right-hand side of (23) can in principle become arbitrarily small. Note that we always have $SNR_{out}^2 / N_q = Q_S^2 (\Delta x)^n / A^n \leq Q_S^2 \leq 1 / C_n$, and hence the right-hand side of (23) is always smaller or equal to 1, i.e. it is weaker than the Heisenberg uncertainty inequality.

Inequality (23) is related to expressions for the information capacity (limits) that were obtained by Shannon for communication systems and by Gabor and others for imaging and electromagnetic fields. According to C. Shannon [12], the number of bits, $N_{bits}$, that can be transmitted within a time interval $A_t$ over a communication channel with bandwidth $W_t = 1 / \Delta t$ is limited by

$$N_{bits} \leq A_t W_t \log SNR. \tag{24}$$

In a related result, Felgett & Linfoot [15] showed that the information capacity of a 2D (incoherent) optical system with the field of view $A_x A_y$ and the spatial bandwidth $W_x W_y = 1/(\Delta x \Delta y)$ is limited by

$$N_{bits} \leq 2 A_x W_x A_y W_y \log SNR. \tag{25}$$

These results were generalized further in [16].

Returning to (23), let $W_{out} = 1 / \Delta x$ be the effective output bandwidth and $A_{out} = 1 / \Delta u$ be the effective output aperture of the imaging system, then (23) can be re-written as

$$SNR_{out}^2 (A_{out} W_{out})^n \leq N_q / C_n. \tag{26}$$

It is easy to see that this is quite a natural inequality, as it states that:
(a) the maximum "information capacity" of an imaging system is limited ultimately by the number of quanta used in the image formation;
(b) the size of an image, its bandwidth and the SNR (or, equivalently, the spatial and angular resolutions, and the SNR) can be traded-off between them, but the product of the three cannot exceed the number of image-forming quanta.

Inequality (26) in 1D and 2D cases (noise - resolution uncertainty) gives complementary results to (24) and (25). Indeed it follows from (26) that $SNR^2 A_x W_x \leq C_1^{-1} N_q$ and $SNR^2 A_x W_x A_y W_y \leq C_2^{-1} N_q$, where we dropped the subscript "out" for brevity. Noting that $(A_{out} W_{out})^n = (A_{out} / \Delta x)^n = N_v$ represents the number of effective resolution units ("voxels"), we obtain in particular that the information capacity of a communication channel or an imaging system in any dimension is ultimately limited by the number of imaging (signal) quanta (e.g. photons) used:

$$N_{bits} \leq N_v \log SNR^2 \leq N_v SNR^2 \leq N_v N_q / (C_n N_v) \leq C_n^{-1} N_q. \tag{27}$$

It will be interesting in the future to consider this result in the context of scattering theory and, in particular, with respect to the limits for the information about the scatterer that can be obtained in a particular imaging scheme (e.g. Computed Tomography) involving a fixed number of scattering (imaging) quanta (e.g. photons). Such an investigation would be relevant to some important practical questions, e.g. those naturally arising in radiation dose limited biomedical imaging or in certain astronomical problems.

## 5. Conclusions

We have derived several forms of uncertainty inequalities which are related to the Heisenberg uncertainty principle and the information capacity of communication and imaging systems described by Shannon and others. We have showed that one of our result (inequality (14)) potentially provides a more accurate lower bound for the phase-space volume (which quantifies the Heisenberg uncertainty) than the conventional uncertainty relationship (7). The new lower bound is related to the "phase-space noise-to-signal ratio" (15) of a given imaging/measuring system. In another result, we suggested an alternative derivation of an "incoherent" version (20) of the Heisenberg uncertainty inequality (which may be termed the Laue inequality [14]). Finally, we obtained an estimate for the information capacity of imaging (scattering) systems which appears complementary to the previous results by Shannon, Gabor and others about the information capacity of communication and imaging systems. According to this last result, the number of bits of information about the sample that can be obtained in an imaging (scattering) experiment cannot exceed the total number of the imaging (scattering) quanta (particles) used in the experiment, while the spatial resolution (number of effective voxels) and the signal-to-noise ratio may be traded for each other.

**References**

[1] M. Born and E. Wolf. *Principles of optics*. Cambridge University Press, Cambridge, 1999.

[2] H.H. Barrett and K.J. Myers. *Foundations of image science*. John Wiley & Sons Inc., Hoboken, 2004.

[3] M. R. Howells, T. Beetz, H. N. Chapman, C. Cui, J. M. Holton, C. J. Jacobsen, J. Kirz, E. Lima, S. Marchesini, H. Miao, D. Sayre, D. A. Shapiro, J. C. H. Spence, and D. Starodub. An assessment of the resolution limitation due to radiation-damage in x-ray diffraction microscopy. *J.Electron Spectros.Relat.Phenomena*, **170**(1-3), 2009, 4–12.

[4] A. C. Fabian, K. A. Pounds and R. D. Blandford. *Frontiers of x-ray astronomy*. Cambridge University Press, Cambridge, 2004.

[5] R. Bach, D. Pope, S.-H. Liou, and H. Batelaan. Controlled double-slit electron diffraction. *New J. Phys.,* **15**, 2013, 033018.

[6] T.E. Gureyev, Ya.I. Nesterets, F. de Hoog, G. Schmalz, S.C. Mayo, S. Mohammadi, and G. Tromba. Duality between noise and spatial resolution in linear systems. *Opt.Express*, **22**(8), 2014, 9087-9094.

[7] T E Gureyev, S C Mayo, Ya I Nesterets, S Mohammadi, D Lockie, R H Menk, F Arfelli, K M Pavlov, M J Kitchen, F Zanconati, C Dullin and G Tromba. Investigation of imaging quality of synchrotron-based phase-contrast mammographic tomography. *J.Phys.D: Appl.Phys.*, **47**(36), 2014, 365401.

[8] F. de Hoog, G. Schmalz, T.E. Gureyev. An uncertainty inequality. *Appl.Math.Letters*, **38**, 2014, 84-86.

[9] Ya.I. Nesterets and T.E. Gureyev. Young's double-slit experiment: noise-resolution duality. *Optics Express*, **23**(3), 2015, 3373–3381.

[10] V.A. Epanechnikov. Non-parametric estimation of a multivariate probability density. *Theory Probab.Appl.*, **14**(1), 1969, 153–158.

[11] G.B. Folland and A. Sitaram. The uncertainty principle: a mathematical survey. *J.Four.Anal.Applic.*, **3**(3), 1997, 207-238.

[12] C. Shannon. Communication in the presence of noise. *Proc.Inst.Radio Engrs.* **37**, 1949, 10-21.

[13] E. Laeng and C. Morpurgo. An uncertainty inequality involving $L_1$ norms. *Proc.Am.Math.Soc.* **127**(12), 1999, 3565-3572.

[14] I. Dreier, W. Ehm, T. Gneiting and D. Richards. Improved bounds for Laue's constant and multivariate extensions. *Math.Nachr.*, **228**, 2001, 109-122.

[15] P.B. Felgett and E.H. Linfoot. On the assessment of optical images. *Phil.Trans.Roy.Soc.A*, **247**, 1955, 369-407.

[16] I. J. Cox, C. J. R. Sheppard. Information capacity and resolution in an optical system. *J.Opt.Soc.Am.A*, **3**, 1986, 1152-1158.

[17] M.G. Cowling and J.F. Price. Bandwidth versus time concentration: the Heisenberg-Pauli-Weyl inequality. *SIAM J.Math.Anal*., **15**(1), 1984, 151-165.

[18] H. Bateman and R. Erdelyi. *Tables of integral transforms*. McGraw-Hill, New York, 1954.

**Appendix**

**Unboundedness of the functional V[t]**

Recall that

$$V\left[t\right] = \frac{\left\|t\right\|_4^4 \left\|\hat{t}\right\|_4^4}{\left\|t\right\|_2^8}.$$

Let us assume that $V\left[t\right] \geq K^8$ for some $K > 0$. From Lemma 3.3 in Folland and Sitaram [11], it follows that this is equivalent to the inequality

$$\left\|t\right\|_4 + \left\|\hat{t}\right\|_4 \geq 2K\left\|t\right\|_2.$$

However, it is shown in Cowling and Price [17] that this inequality is not valid. Thus a non-zero lower bound does not exist.

Turning now to the upper bound, let

$$t\left(x\right) = \left(1 + x^2\right)^{-\frac{3}{8}}$$

for which it is clearly the case that both $\left\|t\right\|_2$ and $\left\|t\right\|_4$ are strictly positive and finite. From Bateman and Erdelyi [18]

$$\hat{t}\left(\xi\right) = \frac{2\pi^{\frac{3}{8}} K_{\frac{1}{8}}\left(2\pi\left|\xi\right|\right)}{\left|\xi\right|^{\frac{1}{8}} \Gamma\left(\frac{3}{8}\right)}$$

where $K_{\nu}$ is a modified Bessel function which has the following behavior for small arguments

$$K_{\nu}\left(z\right) \sim \tfrac{1}{2}\Gamma\left(\nu\right)\left(\tfrac{1}{2}z\right)^{-\nu} \text{ as } z \to 0.$$

Hence

$$\hat{t}^4\left(\xi\right) \sim \left(\frac{\pi^{\frac{1}{4}}\Gamma\left(\frac{1}{8}\right)}{\Gamma\left(\frac{3}{8}\right)}\right)^4 \left|\xi\right|^{-1} \text{ as } \xi \to 0$$

and, as a consequence, $\left\|\hat{t}\right\|_4$ is unbounded. Thus $V\left[t\right]$ does not have a finite upper bound.